\documentclass[proceedings]{JHEP3} % 10pt is ignored!
\usepackage{epsfig,multicol}

\newbox\mybox
\newcommand\fverb{\setbox\mybox=\hbox\bgroup\verb}
\newcommand\fverbdo{\egroup\medskip\noindent\fbox{\unhbox\mybox}\ }
\newcommand\fverbit{\egroup\item[\fbox{\unhbox\mybox}]}

\title{CP Violation}

\author{M. Beneke\thanks{Talk presented at the International 
        Europhysics Conference on High Energy Physics, Budapest, 
        July 2001.}\\
        Institut f\"ur Theoretische Physik E, Sommerfeldstr. 28,\\ 
        RWTH Aachen, D -  52074 Aachen, Germany}
       
\conference{PITHA 02/03, hep-ph/0201137, 15 January 2002}

\abstract{
Several pieces of direct and indirect evidence now suggest 
that the Kobayashi-Maskawa mechanism plays a distinguished 
role for CP violation at the electroweak scale. This 
talk provides a general overview of CP violation in its 
various contexts, emphasizing CP violation in flavour-violating 
interactions, such as due to the Kobayashi-Maskawa mechanism. 
I then review a few recent theoretical developments relevant 
to the interpretation of CP violation.\footnote{CP violation is 
inseparably linked to flavour physics in the Standard Model. Three 
plenary talks have been devoted to this subject at this conference. 
The talk by Ligeti \cite{Ligeti:2001an} reports on recent results 
in flavour physics excluding CP violation. The talk by Hamel de 
Monchenault \cite{HameldeMonchenault:2001ae} summarizes experimental 
results on CP violation in $B$ decays. CP violation is currently a 
rather phenomenological subject and some overlap of this talk with 
\cite{HameldeMonchenault:2001ae} is unavoidable. In addition to 
discussing theoretical aspects of CP violation, this talk  
also covers recent experimental results on CP violation in the kaon 
system.}
}

\begin{document} 

\section{Introduction}

The breaking of CP symmetry (``CP violation''), the composition of 
parity and charge conjugation, is an interesting phenomenon for 
several reasons:

a. CP violation together with CPT symmetry implies non-invariance of
the microscopic equations of motion under time-reversal. CP violation 
rather than C violation implies different physical properties of
matter and antimatter. These two facts are of fundamental importance 
for our understanding of the laws of Nature, and they were perceived
as  revolutionary, when CP violation was discovered in 
1964 \cite{Christenson:fg}. They were also important in the
development of the 
fundamental theory of particles, since the observation of CP violation 
motivated some early extensions of the 
Standard Model as it was known at the time (1973), either by extending 
the Higgs sector \cite{Lee:iz} or by adding a third generation of 
quarks and leptons \cite{Kobayashi:fv}. 
Nature has opted for the second possibility  
for certain, and the Kobayashi-Maskawa mechanism of CP violation has become 
part of today's Standard Model. From today's perspective
time-reversal non-invariance and the distinction of matter and 
antimatter, though fundamental, appear no longer surprising and even 
``natural''. What remains surprising, however, is the peculiar way 
in which CP violation occurs or rather does not occur in the Standard Model 
and its possible extensions.

b. Assuming that the evolution of the universe began from a
matter-antimatter symmetric state, CP violation is necessary 
\cite{Sakharov:dj} to 
generate the matter-antimatter asymmetric state of the 
universe that one observes
today. At the electroweak phase transition the Standard Model 
satisfies all the other necessary criteria for baryogenesis (baryon 
number violation, departure from thermal equilibrium) 
\cite{Kuzmin:1985mm}, but CP
violation in the Standard Model is too weak to explain the observed
baryon-to-photon ratio. With the above assumption on the initial 
condition of the cosmic evolution, our own existence provides evidence
for a source of CP violation beyond the Standard Model.

Electroweak baryogenesis has the attractive feature that it couples 
the required new mechanisms of CP violation to the electroweak scale, 
therefore making them testable also in particle collider experiments. 
Nevertheless it now appears more likely that the matter-antimatter
asymmetry is not related to the sources of CP violation that one may 
observe at colliders. Two facts have contributed to this change of
perspective: first, the lower limit on the masses of Higgs bosons has
been increasing. A heavier Higgs boson implies a weaker (first-order)
electroweak phase transition. As a consequence electroweak
baryogenesis is already too weak over most of the parameter space of
even the minimal supersymmetric extension of the Standard Model. Second, 
the observation of small neutrino masses through neutrino oscillations 
is explained most naturally by invoking the seesaw mechanism, 
which in turn is most naturally realized by postulating
massive neutrinos, which are singlets under the Standard Model gauge
group. All three necessary conditions for the generation of lepton
number are naturally realized in the decay of the massive neutrino(s). 
Lepton number is then partially converted into baryon number via 
B+L-violating (but B--L conserving) sphaleron transitions 
\cite{Fukugita:1986hr}. While the 
leptogenesis scenario is very appealing, the new sources of CP
violation related to the Yukawa couplings of the heavy neutrinos occur 
at scales of order of the heavy neutrino mass, $M_R\sim
(10^{12}-10^{16})\,\mbox{GeV}$ (needed to explain the small left-handed 
neutrino masses), and are not directly testable with collider
experiments in the near future. 
For this reason, CP violation in the context of baryogenesis will not 
be discussed further in this talk.

c. CP violation in the Standard Model is essentially an electroweak 
phenomenon originating from the Yukawa couplings of the quarks to the 
Higgs boson. 
This implies that probes of CP violation are indirect 
probes of the electroweak scale or TeV scale, complementary to direct
probes such as the observation of Higgs bosons. This is probably the 
most important reason for the current interest in CP symmetry 
breaking: in addition to testing the Kobayashi-Maskawa mechanism
of CP violation in the Standard Model, experiments directed at CP
violation limit the construction of extensions of the Standard Model
at the TeV scale. There is an analogy between CP symmetry and 
electroweak symmetry breaking. Both occur at the electroweak scale 
and for both the Standard Model provides a simple mechanism. However,
neither of the two symmetry breaking mechanisms has been sufficiently 
tested up to now. 

d. Leaving aside the matter-antimatter asymmetry in the universe as
evidence for CP violation since this depends on a further assumption, 
CP violation has now been observed in the weak interactions of quarks 
in three different ways: in the mixing of the neutral kaon flavour 
eigenstates ($\epsilon$, 1964) \cite{Christenson:fg}; 
in the decay amplitudes of neutral 
kaons ($\epsilon^\prime/\epsilon$, 1999) 
\cite{Alavi-Harati:1999xp,Fanti:1999nm}; in the mixing of the neutral
$B_d$ meson flavour eigenstates ($\sin(2\beta)$, 2001) 
\cite{Aubert:2001nu,Abe:2001xe}. It will be 
seen below that these pieces of data together with others not directly
related to CP violation suggest that the Kobayashi-Maskawa mechanism
of CP violation is most likely the dominant source of CP violation at
the electroweak scale. The latest piece of evidence also rules out
that CP symmetry is an approximate symmetry. As a consequence 
generic extensions of the Standard Model at the TeV scale needed
to explain the stability of the electroweak scale suffer from a CP
fine-tuning problem since any such extension implies the existence of
many new CP-violating parameters which have no generic reason to be
small, but which put these theories into conflict with experiment, if
they are large. Despite the apparent success of the standard theory of CP
violation, the problem of CP and flavor violation therefore remains as
mysterious as before. 

\section{CP violation in the Standard Model}

CP violation can occur in the Standard Model in three different ways:

\subsection{The $\theta$ term}

The strong interactions could be CP-violating 
\cite{'tHooft:up,Jackiw:1976pf,Callan:je}. The topology of gauge
fields implies that the correct vacuum 
is given by a superposition
$|\theta\rangle = \sum_n e^{i n\theta} |n\rangle$
%$|\theta\rangle = \sum_{n=-\infty}^\infty e^{i n\theta} |n\rangle$
%\begin{equation}
%|\theta\rangle = \sum_{n=-\infty}^\infty e^{i n\theta} |n\rangle
%\end{equation}
of the degenerate vacua $|n\rangle$ in which pure gauge fields have 
winding number $n$. Correlation functions in the $\theta$-vacuum can 
be computed by adding to the Lagrangian the term
\begin{equation}
{\cal L}_\theta = \theta\cdot\frac{g_s^2}{32\pi^2}\,G_{\mu\nu}^A 
\tilde G^{A,\mu\nu},
\end{equation}
where $\theta$ now represents a parameter of the theory. Physical
observables can depend on $\theta$ only through the combination 
$e^{i\theta} \det {\cal M}$, where ${\cal M}$ is the quark mass
matrix. A non-zero value of 
$\tilde \theta =\theta + \mbox{arg}\det {\cal M}$ 
%\begin{equation}
%\tilde \theta =\theta + \mbox{arg}\det {\cal M}
%\end{equation}
violates CP symmetry. It also implies an electric dipole moment of 
the neutron of order $10^{-16}\,\tilde \theta \,e\,\mbox{cm}$ 
\cite{Baluni:1978rf}. The 
non-observation of any such electric dipole moment constrains 
$\tilde \theta < 10^{-10}$ and causes what is known as the strong CP problem, 
since the Standard Model provides no mechanism that would require 
$\tilde \theta$ to vanish naturally. The strong CP problem has become 
more severe with the observation of large CP violation in $B$ meson
decays, since one now knows with more confidence that the quark mass 
matrix has no reason to be real a priori. 

There exist mechanisms that render $\tilde\theta$ exactly zero or very 
small through renormalization effects. None of these mechanisms is 
convincing enough to provide a default solution to the problem. 
What makes the strong CP problem so difficult to solve is that one
does not have a clue at what energy scale the solution should be 
sought. Strong CP violation is not discussed further in this talk  
(see the discussion in \cite{Peccei:1998jt}).

\subsection{The neutrino mass matrix}

The Standard Model is an effective theory defined by its gauge 
symmetries and its particle content. CP violation appears in the
lepton sector if neutrinos are massive. The leading operator 
in the effective Lagrangian is \cite{Weinberg:sa}
\begin{equation}
{\displaystyle \frac{f_{ij}}{\Lambda}}
\cdot [(L^T\epsilon)_i i\sigma^2 H][H^T i\sigma^2
L_j],
\end{equation}
where $L_i$ and $H$ denote the lepton and Higgs doublets,
respectively. 
After electroweak symmetry breaking this generates a Majorana neutrino
mass matrix with three CP-violating phases. One of these phases could
be observed in neutrino oscillations, the other two phases only in 
observables sensitive to the Majorana nature of neutrinos.

Unless the couplings $f_{ij}$ are extremely small, the scale $\Lambda$ must be
large to account for small neutrinos masses, which suggests that
leptonic CP violation is related to very large scales. For example, 
the standard see-saw mechanism makes the $f_{ij}$ dependent on the 
CP-violating phases in the heavy gauge-singlet neutrino mass matrix. 
As a consequence one may have interesting model-dependent relations 
between leptogenesis, CP violation in lepton-flavour violating
processes and neutrino physics, but since the observations are all
indirect through low-energy experiments, one may at best hope for
accumulating enough evidence to make a particular model particularly
plausible. Such experiments seem to be possible, but not in the near
future, and for this reason leptonic CP violation is not discussed
further here. It should be noted that there is in general no connection 
between CP violation in the quark and lepton sector except in grand
unification models, where the two relevant Yukawa matrices are
related. Even then, further assumptions are necessary for a
quantitative relation. 

\subsection{The CKM matrix}

CP violation can appear in the quark sector of the Standard Model 
at the level of renormalizable interactions \cite{Kobayashi:fv}. 
The quark Yukawa
interactions read
\begin{equation}
{\cal L}_Y = -y^d_{ij} \bar Q^\prime_i H d^\prime_{Rj} - 
y^u_{ij} \bar Q^\prime_i \epsilon H^*u^\prime_{Rj} + {\rm h.c.},
\end{equation}
with $Q^\prime $ the left-handed quark SU(2)-doublets, $u_R^\prime$ and 
$d_R^\prime$ the right-handed SU(2)-singlets and $i,j=1,2,3$ 
generation indeces. The complex mass matrices
that arise after electroweak symmetry breaking are diagonalized by 
separate unitary transformations $U^{u,d}_{L,R}$ of the left- and
right-handed up- and down-type fields. Only the combination 
\begin{equation}
V_{\rm CKM} = {U_L^u}^\dagger U_L^d = 
\left(\begin{array}{ccc}
V_{ud} & V_{us} & V_{ub}\\
V_{cd} & V_{cs} & V_{cb}\\
V_{td} & V_{ts} & V_{tb}
\end{array}\right),
\end{equation}
referred to as the CKM matrix, is observable, 
since the charged current interactions now read
\begin{equation}
-\frac{e}{\sqrt{2}\sin\theta_W} \bar u_{Li} \gamma^\mu [V_{\rm CKM}]_{ij}
d_{Lj} W^+_\mu + \mbox{h.c.}.
\end{equation}
At tree level flavour and CP violation in the quark sector 
can occur in the Standard Model only through charged current 
interactions (assuming $\tilde\theta =0$). With three generations of
quarks, the CKM matrix contains one physical CP-violating phase. 
Any CP-violating observable in flavour-violating processes must be
related to this single phase. The verification or, perhaps rather,
falsification of this highly
constrained scenario is the primary goal of many current 
$B$- and $K$-physics experiments. This type of CP violation is
therefore discussed in some detail in later sections.

For reasons not understood the CKM matrix has a hierarchical structure
as regards transitions between generations. It is therefore often
represented in the approximate form \cite{Wolfenstein:1983yz}
\arraycolsep0.05cm
\begin{equation}
V^{\rm CKM} = 
\left(
\begin{array}{ccc}
1-\lambda^2/2 & \lambda & 
A\lambda^3 (\rho-i\eta) \\[0.2cm]
-\lambda & 1-\lambda^2/2 & A\lambda^2 \\[0.2cm]
A\lambda^3 (1-\rho-i\eta) & -A\lambda^2 & 1
\end{array}\right) 
+ O(\lambda^4)
,
\end{equation}
where $\lambda\approx 0.224$ and $A$, $\rho$, $\eta$ are 
counted as order unity. 
%and corrections are of order $\lambda^4$. 
%It is the great achievement of heavy quark theory of the 1990s
%to have determined $|V_{cb}|$, i.e. $A$, to the accuracy of a few
%percent, whereas determining $\rho$ and $\eta$ with this accuracy 
%remains a challenge for this decade. 
The unitarity of the CKM matrix 
leads to a number of relations between
rows and columns of the matrix. The one
which is most useful for $B$-physics is obtained by multiplying the
first column by the complex conjugate of the third:
\begin{equation}
V_{ud} V_{ub}^*+V_{cd} V_{cb}^*+V_{td} V_{tb}^*=0.
\end{equation}
If $\eta\not=0$ (which implies CP violation) this relation can be
represented as a triangle in the complex plane, called the unitarity
triangle. See Figure~\ref{fig1}, which also introduces some notation
for the angles of the triangle that will be referred to later on.
\FIGURE{
\epsfig{file=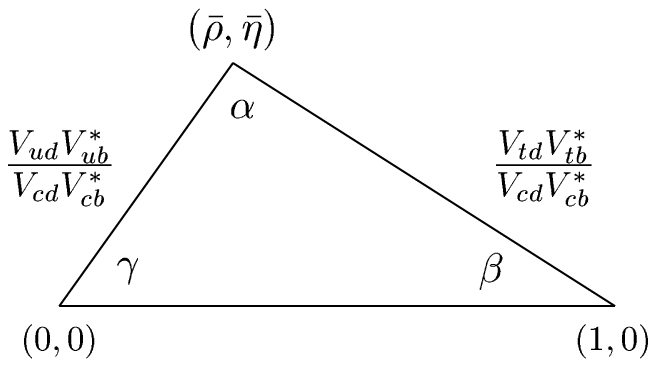, width=.39\textwidth}
\caption{The unitarity triangle.}
\label{fig1}}

%%%%%%%%%%%%%%%%%%%%%%%%%%%%%%%%%%%%%%%%%%%%%%%%%%%%%%%%%%%%%%%%%%%
%\begin{figure}[t]
%   \vspace{-2.3cm}
%   \epsfysize=24cm
%   \epsfxsize=16cm
%   \hspace*{-1cm}
%   \epsffile{ut.ps}
%   \vspace*{-19.6cm}
%\caption[dummy]{\label{fig1} The unitarity triangle.}
%\end{figure}
%%%%%%%%%%%%%%%%%%%%%%%%%%%%%%%%%%%%%%%%%%%%%%%%%%%%%%%%%%%%%%%%%%%
The hierarchy of the CKM matrix implies that CP violation is a small
effect in the Standard Model even when the CP-violating phase is
large. More precisely, CP-violating observables
are either small numbers, or else they are constructed out of small
numbers such as small branching fractions of rare decays. The
hierarchy of quark mas\-ses and mixing angles represents a puzzle, 
sometimes called the flavour problem, which will also not be discussed
further in this talk. 
%Since the relevant couplings are all
%renormalizable, the Standard Model itself can offer no insight into
%this problem. 
%Typical attempts to solve the flavour problem  
%focus on broken generation symmetries.

\section{Kaon decays}

Rare kaon decays have been very important for the construction of 
the Standard Model and for CP violation in particular. 
Kaons continue to be among the most sensitive probes of flavour-changing 
interactions. This is illustrated by the suppression of $|\Delta S|=2$
transitions in the Standard Model due to weak coupling, the GIM
mechanism and small CKM matrix elements. None of these suppressions 
needs to hold in extensions of the Standard Model. 

\subsection{CP violation in mixing (indirect)} Due to CP violation 
in $K\bar K$ mixing, the neutral kaon mass eigenstates are superpositions 
of CP-even and CP-odd components. The long-lived kaon state, 
$K_L \approx K_2+\bar\epsilon K_1$, is predominantly CP-odd, but decays 
into two pions through its small CP-even component $K_1$. The decay 
$K_L\to\pi\pi$ constituted the first observation of CP violation 
ever \cite{Christenson:fg}. 
The quantity $|\epsilon|=2.27\cdot 10^{-3}$ (equal to $|\bar\epsilon|$ 
in the standard phase convention to very good accuracy) 
has now been measured in many different ways. There are two 
new results related to $\bar \epsilon$: 

\begin{itemize}
\item[1.] The charge asymmetry in $K_L\to \pi e \nu$ decay has been measured 
very precisely by KTeV \cite{2ktev2001}:
\begin{equation}
2\,\mbox{Re} \,{\bar\epsilon} 
\approx \frac{\Gamma(\pi^- e^+\nu)-\Gamma(\pi^+ e^-\bar\nu)}
{\Gamma(\pi^- e^+\nu)+\Gamma(\pi^+ e^-\bar\nu)} 
= (3.322\pm 0.074)\cdot 10^{-3}.
\end{equation}

\item[2.] The rare decay $K_L\to \pi^+\pi^- e^+ e^-$ can proceed through the 
CP-violating $K_L\to\pi^+\pi^-$ decay with subsequent radiation of 
a virtual photon which converts into an $e^+ e^-$-pair. The amplitude 
for this decay is proportional to $\bar \epsilon$. It can also 
proceed through a CP-conserving direct 
amplitude $K_L\to \pi^+\pi^-\gamma^*$. Since this amplitude is 
dynamically suppressed, there exists the possibility of a large 
CP-violating asymmetry. KTeV and NA48 investigated the angular 
distribution $d\Gamma/d\phi$,
where $\phi$ is the angle between the $\pi^+\pi^-$ and the $e^+ e^-$ 
decay planes. The decay plane asymmetry is CP-violating 
and has been found to be \cite{na482001,Alav-Harati:1999ff}
\begin{equation}
A\equiv \left|\frac{1}{\Gamma}
\int d\phi\,\frac{d\Gamma}{d\phi}\mbox{
sign}(\sin(2\phi))\right|\\[0.3cm]
=
\left\{\begin{array}{ll}
(13.9\pm 2.7\pm 2.0)\% & \quad \mbox{NA48}\\
(13.6\pm 2.5\pm 1.2)\% & \quad \mbox{KTeV}
\end{array}\right.,
\end{equation}
in agreement with theoretical expectations \cite{Sehgal:wm}.
\end{itemize}

\subsection{CP violation in decay (direct)} 

CP-violating effects in kaon decays can also occur due to 
the interference of two decay amplitudes with different CP-violating 
phases independent of CP violation in $K\bar K$ mixing. In the 
Standard Model both manifestations of CP violation are related, 
but one can construct models in which CP violation arises only 
in $|\Delta S|=2$ interactions and not in the decay amplitude. 

There are a number of searches for direct CP violation with no result 
expected in the Standard Model at the current sensitivities of the 
experiments. The HyperCP experiment searches for 
direct CP violation in hyperon decays and in charged kaon decays to 
three pions. The CP-violating asymmetries in hyperon decay are
compatible with zero with an error of about $2\cdot 10^{-3}$ 
\cite{Leros:hz,dukes} compared to a Standard Model expectation of 
$10^{-(4-5)}$. Another example is the transverse muon polarization 
\begin{equation}
P_\perp = \langle \vec{S}_\mu\cdot \frac{(\vec p_\pi\times \vec p_\mu)}
{|\vec p_\pi\times \vec p_\mu|}\rangle 
\end{equation}
in $K^+\to \pi^0 \mu^+\nu$ decay. 
This observable is odd under naive time-reversal (no exchange of
initial and final state) and can therefore also be generated by 
(electromagnetic) final state interactions. With a neutral pion 
in the final state this effect is very small \cite{Zhitnitsky:he} 
resulting in $P_\perp\sim 10^{-(6-7)}$ in the Standard Model. 
CP-violating interactions in 
extensions of the Standard Model could yield a transverse polarization
of order $10^{-3}$ through interference with the dominant 
charged current amplitude. The current result 
$P_\perp =  -(3.3\pm 3.7 \pm 0.9)\cdot 10^{-3}$ by the KEK-E246 
experiment \cite{kek246} therefore begins to reach an interesting level of 
sensitivity.

Much experimental effort has been invested into the search for 
direct CP violation in kaon decays to two pions. The double ratio
\begin{equation}
\frac{\Gamma(K_L\to\pi^0\pi^0)\Gamma(K_S\to\pi^+\pi^-)}
{\Gamma(K_S\to\pi^0\pi^0)\Gamma(K_L\to\pi^+\pi^-)}
\approx 1-6\mbox{Re}\left(\frac{\epsilon'}{\epsilon}\right),
\end{equation}
if different from unity, implies such an effect, since both ratios 
would equal $|\epsilon|$, if CP violation occurred only in mixing. 
The existence of this effect has been conclusively demonstrated by 
two experiments in 1999, following the first hints of a non-vanishing 
$\epsilon'/\epsilon$ in 1992 \cite{Barr:1993rx}. 
The new results of 2001 have further 
clarified the situation, which is now summarized by 
\cite{Lai:2001ki,ktev2001}
\begin{eqnarray}
{\displaystyle \frac{\epsilon^\prime}{\epsilon}} =
\left\{\begin{array}{ll}
(15.3 \pm 2.6)\cdot 10^{-4} & \quad\mbox{NA48 (97-99)} \\[0.0cm]
(20.7 \pm 2.8)\cdot 10^{-4} & \quad\mbox{KTeV (96/97)}.
\end{array}\right.
\end{eqnarray}

The theory of $\epsilon'/\epsilon$ is rather complicated. 
The short-distance contributions have been worked out to 
next-to-leading order \cite{Ciuchini:1992tj,Buras:1993dy} and do not 
constitute a major source of uncertainty, but the remaining hadronic 
matrix elements continue to prevent an accurate computation of 
$\epsilon'/\epsilon$. The following approximate representation 
of the result \cite{buras},
\begin{eqnarray}
\frac{\epsilon^\prime}{\epsilon} &=&
16\cdot 10^{-4} \left[{\displaystyle \frac{\mbox{Im}\,V_{ts}^*
      V_{td}}{1.2\cdot 10^{-4}}}\right]
\left({\displaystyle
    \frac{110\,\mbox{MeV}}{m_s(2\,\mbox{GeV})}}\right)^2\\
&& \hspace*{0.0cm}
\times \left\{\underbrace{B_6^{(1/2)} (1-\Omega_{IB})} 
- \underbrace{0.4 \left({\displaystyle
    \frac{\bar m_t}{165\,\mbox{GeV}}}\right)^{2.5}\!B_8^{(3/2)}} 
\right\},\nonumber\\
&&\hspace*{1.3cm}\mbox{QCD penguin}\quad\qquad \mbox{
EW penguin}\nonumber
\end{eqnarray}
illustrates the difficulty that arises from a cancellation between 
strong and electroweak penguin contributions and the need to know 
the hadronic matrix elements $B_i \propto 
\langle \pi \pi |O_i| K\rangle$, which involve a two-pion final 
state, accurately. Before 1999 it was commonly, though not universally, 
assumed that $B_{6,8}\approx 1$ near their vacuum saturation value, 
and with the isospin breaking factor $\Omega_{IB}\approx 0.25$, this 
gives only about $6\cdot 10^{-4}$. The experimental result has triggered 
a large theoretical activity over the past two years 
directed towards better understanding 
the hadronic matrix elements. Alternatively, extensions of the 
Standard Model have been invoked to explain a supposed enhancement of
$\epsilon'/\epsilon$. 

I am not in the position to review this activity in detail, a fraction
of which has been represented at this conference 
\cite{Blum:2001xb,Boucaud:2001mg,Cir:2001qw,Knecht:2001bc,Sci:2001kc}.
Different approaches continue to disagree 
by large factors, but it appears now certain that serious hadronic matrix 
element calculations must account for final 
state interactions of the two pions. Since $\epsilon'/\epsilon$ is 
now accurately known experimentally and could be used to constrain 
$\eta$, the CP-violating parameter of the Standard Model, or 
new CP-violating interactions, it will be important in the 
future that theoretical calculations have control over the
approximations involved. 
Chiral perturbation theory 
combined with a large-$N_c$ matching of the operators in 
the non-leptonic Hamiltonian can probably go furthest towards 
this goal with analytic methods. The calculation reported in 
\cite{Pallante:2001he} finds $B_6^{(1/2)}$ enhanced by a 
factor 1.55 through $\pi\pi$ rescattering in the isoscalar channel, 
while $B_8^{(3/2)}$ remains close to 1. When this is combined with 
a re-evaluation of isospin breaking which gives $\Omega_{IB}$ smaller 
than before \cite{Ecker:1999kr,Gardner:1999wb,Wolfe:2000rf}, 
chiral perturbation theory supplemented by a resummation of
rescattering effects may account for the experimental result within 
theoretical uncertainties. However, the same approach does not 
explain the enhancement of the real part of the $\Delta
I=1/2$-amplitude known as the $\Delta I=1/2$ rule.
As a matter of principle, lattice QCD can settle the 
matrix element calculation definitively, since $K\to \pi\pi$ matrix 
elements computed in a lattice of finite volume, can be matched 
to continuum, infinite-volume matrix elements including all information 
on rescattering \cite{Lellouch:2000pv,Lin:2001ek}. Due to 
the potential cancellations the matrix elements 
will, however, be needed with high precision. 

Despite the fact that $\epsilon'/\epsilon$ cannot be computed with 
great precision at present and converted into a stringent test of the CKM 
structure, the fact that direct CP violation is large is important for
model building. Also, in the Standard Model, $\epsilon'/\epsilon$ 
provides further evidence that $\eta$, i.e. the CKM phase is 
sizeable.

\subsection{Very rare kaon decays} There exist several proposals 
(E949 and KOPIO at Brookhaven; CKM and KAMI at Fermilab) to 
measure the very rare decays 
$K^+\to\pi^+\nu\bar\nu$ and $K_L\to\pi^0\nu\bar\nu$ with 
expected branching fractions of about $(7.5\pm 3)\times 10^{-11}$ and 
$(2.6\pm 1.2)\times 10^{-11}$, respectively. The first of these decays is 
CP-conserving and constrains $(\rho,\eta)$ to lie on a certain 
ellipse in the $(\rho,\eta)$-plane. The second decay is 
CP-violating and determines $\eta$. The branching fractions 
are predicted theoretically with high precision: the short-distance 
corrections are known to next-to-leading order \cite{Buchalla:1998ba} 
and the hadronic matrix elements can be obtained from semi-leptonic
kaon decays. The $K^+\to\pi^+\nu\bar\nu$ decay is predicted less 
accurately than $K_L\to\pi^0\nu\bar\nu$ due to the presence of a charm
contribution. However, potential non-perturbative
$1/m_c^2$-corrections due to this charm contribution have been
estimated to be small \cite{Falk:2000nm}. Therefore these two kaon
modes alone, given precise measurements of these experimentally rather
challenging decays, can fix the shape of the unitarity triangle 
precisely, or uncover inconsistencies with other constraints. 
In fact, the $K\to\pi\nu\bar\nu$ modes are themselves sensitive 
probes of modifications of the effective flavour-changing $\bar s d Z$ 
coupling. Two\footnote{The second event was published after the
  conference.} $K^+\to\pi^+\nu\bar\nu$ events have been 
observed by the BNL E787 experiment \cite{unknown:2001xv} 
resulting in a branching fraction 
$\mbox{Br}(K^+\to \pi^+\nu\bar\nu) = (16^{+18}_{-8})\times 10^{-11}$, 
somewhat larger than 
expected but consistent with expectations within the experimental error.

\section{Constraints on the unitarity triangle} 

The CP-violating quantity $\epsilon$ in $K\bar K$ mixing,
$|V_{ub}/V_{cb}|$, and the mass differences of the neutral $B$ mesons, 
$\Delta M_{B_d}$, $\Delta M_{B_s}$, are used to constrain $(\bar \rho,
\bar\eta)$, the apex of the unitarity 
triangle.\footnote{The definitions $\bar\rho = 
\rho\,(1 -\lambda^2/2)$, $\bar\eta = 
\eta\,(1 -\lambda^2/2)$ render the location of the apex 
accurate to order $\lambda^5$ \cite{Buras:1994ec} 
and will be used in the following.} I take here the point of view 
that this constrains the quantity $\sin (2\beta)$ indirectly. The
range obtained is then compared with the direct measurement of 
$\sin (2\beta)$ (discussed in the next section). Alternatively, 
one could include the direct measurement of 
$\sin (2\beta)$ as a fifth observable into the fit. 
It is not the purpose of this talk to go into the details of the 
theoretical calculations that 
contribute to these constraints, since this would lead away from the
topic of CP violation.  Recent summaries of  
lattice calculations of the relevant hadronic parameters can be found in 
\cite{Ryan:2001ej,Martinelli:2001yn,Kronfeld:2001ss}. The status 
of the determination of $|V_{cb}|$ and $|V_{ub}|$ is reviewed in 
the talk by Ligeti \cite{Ligeti:2001an}. 

The following equations summarize the four constraints in compact 
form. With $\lambda=0.224$, $|V_{cb}|=0.041\pm 0.002$, 
$\epsilon=(2.280\pm 0.019)\times 10^{-3}$ and 
$\Delta M_{B_d}=(0.487\pm 0.014)\,\mbox{ps}^{-1}$, and neglecting 
small errors, one obtains:

\TABLE{
\begin{tabular}{c|c|c} \hline
Observable & Constraint  &  Dominant error \\ \hline\hline
&&\\[-0.35cm]
${\displaystyle \left|\frac{V_{ub}}{V_{cb}}\right|}$ & 
$\sqrt{\bar \rho^2+\bar\eta^2} = 0.37\times
{\displaystyle \frac{|V_{ub}/V_{cb}|}{0.085}}$&
$\pm20\%$ ($|V_{ub}|$) \\
$\Delta M_{B_d}$ & 
$\sqrt{(1-\bar\rho)^2+\bar\eta^2} = 0.83 \times 
{\displaystyle \frac{f_{B_d} B_{B_d}^{1/2}}{230\,{\rm MeV}}}$ & 
$\pm 15\%$ ($f_{B_d} B_{B_d}^{1/2}$) \\
&&\\[-0.35cm]
${\displaystyle \frac{\Delta M_{B_d}}{\Delta M_{B_s}}}$ &
$\sqrt{(1-\bar\rho)^2+\bar\eta^2} = 0.87\times
{\displaystyle \frac{\xi}{1.16}}\times
\sqrt{{\displaystyle 17.5\,\mbox{ps}^{-1}/\Delta M_{B_s}}}$& 
$\pm 6\%$ ($\xi$)\\
&&\\[-0.4cm]
$\epsilon$ &
$\bar\eta \,(1.31\pm 0.05-\bar\rho) = 
0.35\times {\displaystyle 0.87/\hat B_K}$ 
& $\pm15\%$ ($\hat B_K$) \\[-0.4cm] 
&&\\ \hline
\end{tabular}
}

\noindent 
One notes that the dominant uncertainties are theoretical except,
perhaps, for $|V_{ub}|$, for which the relative size of 
experimental and theoretical errors depends on the method of
determination. Also the constraint from $\Delta M_{B_s}$ currently 
gives only an upper bound on one of the sides of the unitarity
triangle, since only a lower bound on $\Delta M_{B_s}$ is measured. 

It is in principle straightforward to fit $(\bar \rho,
\bar\eta)$ to the four observables listed above except for the 
fact that the dominant errors are theoretical and therefore do not 
(usually) admit a statistical interpretation. Different statistical 
procedures are being used (``frequentist'', ``Bayesian'', ``scanning'',
``Gaussian'', etc.). Figure~\ref{fig2} shows two representatives of
such global fits, one using a variant of the scanning approach  
\cite{Hocker:2001xe} (upper panel), 
the other using the Bayesian (or inferential) 
approach \cite{Ciuchini:2000de} (lower panel).

%%%%%%%%%%%%%%%%%%%%%%%%%%%%%%%%%%%%%%%%%%%%%%%%%%%%%%%%%%%%%%%%%%%
\begin{figure}[t]
%\vspace*{0.8cm}
\hspace*{2.3cm}\epsfig{file=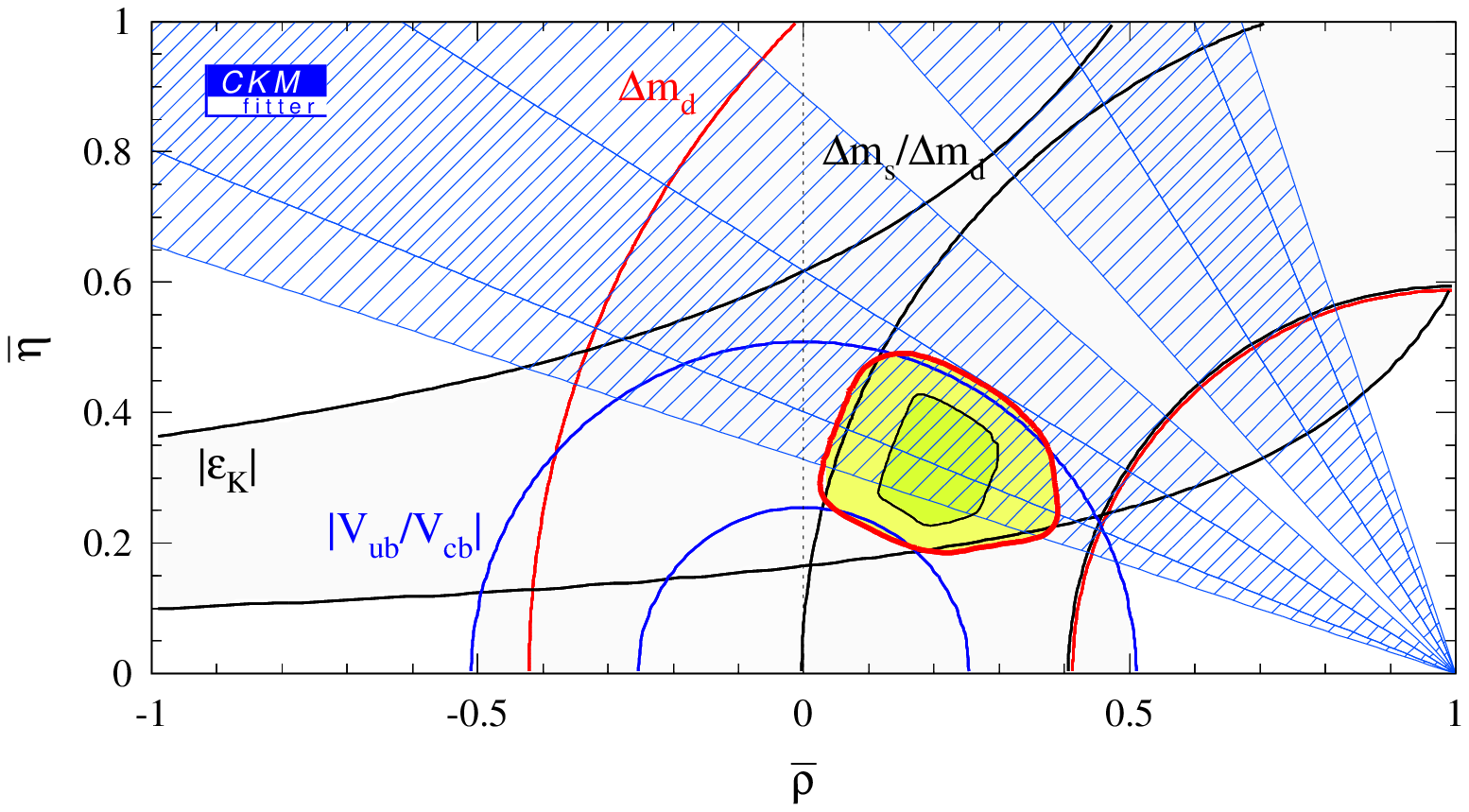, width=.63\textwidth}
\end{figure}
%\vskip-2cm
\begin{figure}[t]
\vskip-1.2cm
\hspace*{1.6cm}\epsfig{file=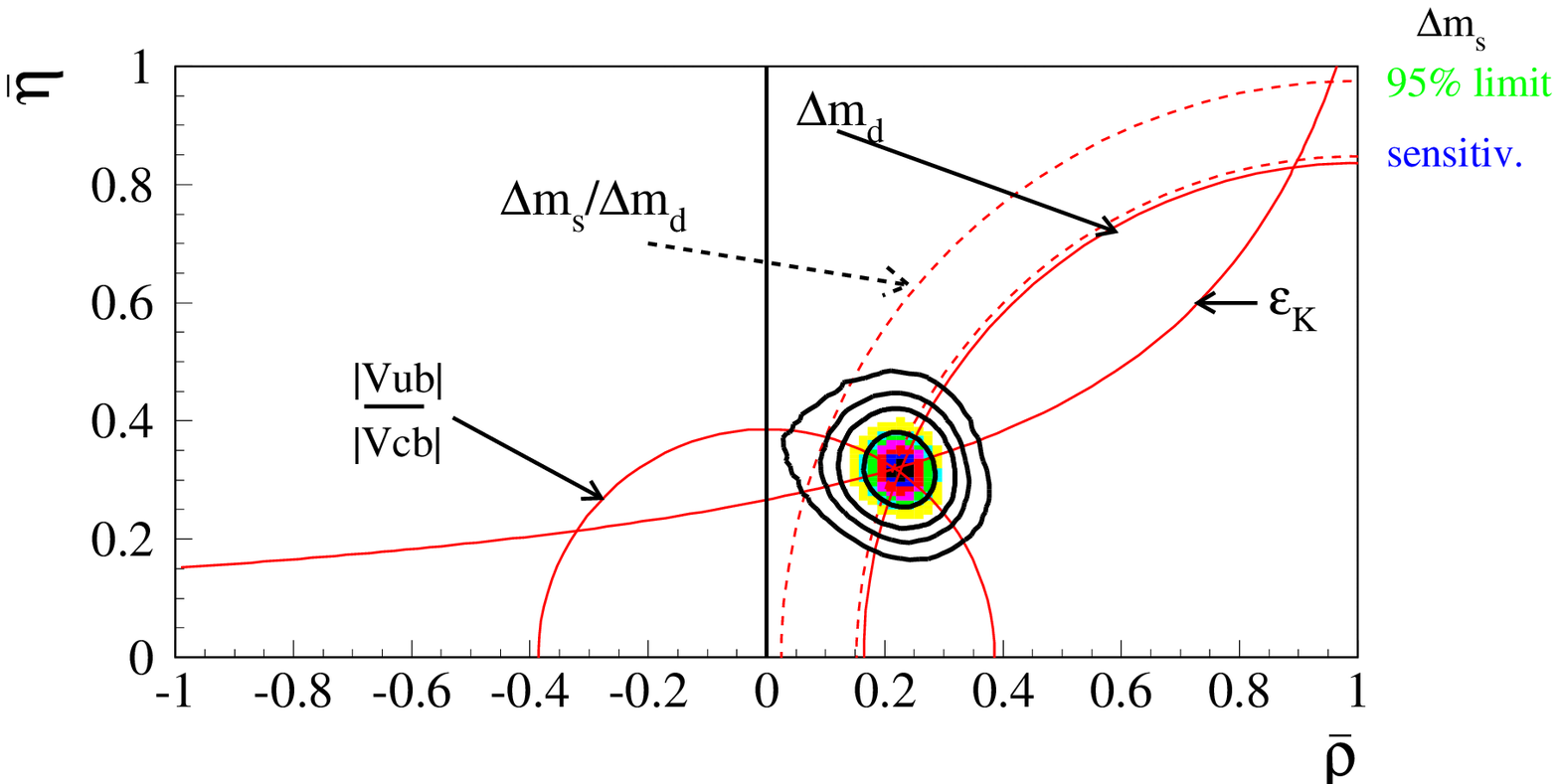, width=.78\textwidth}
\caption[dummy]{\label{fig2} Two summaries 
of unitarity triangle constraints 
(excluding the direct measurement of $\sin(2\beta)$ which is overlaid 
in the upper panel) \cite{Hocker:2001xe,Ciuchini:2000de}.}
\end{figure}
%%%%%%%%%%%%%%%%%%%%%%%%%%%%%%%%%%%%%%%%%%%%%%%%%%%%%%%%%%%%%%%%%%%

The four quantities are seen to be in remarkable agreement and 
reveal no sign of inconsistency of the Kobayashi-Maskawa mechanism for 
CP violation. Also the various statistical procedures appear to give
similar results when the same inputs are used (which is not strictly 
the case in the Figure above). The combined fit results 
in $\bar\rho = 0.21\pm 0.17$, $\bar\eta = 0.35\pm 0.14$ and 
an indirect determination of the angles of the unitarity triangle,
\begin{equation}
\label{angles}
\sin(2\alpha) = -0.24\pm 0.72,\quad
\sin(2\beta)=0.68\pm 0.21,\quad
\gamma=(58\pm24)^\circ,
\end{equation}
where the errors should imply a $95\%$ confidence level \cite{Hocker:2001xe}.
%, even though 
%the statistical interpretation of this is not obvious.

In the future one can expect improvements in this fit from a better 
measurement of $|V_{ub}|$ and from progress in lattice QCD 
on the relevant quantities. If $\Delta M_{B_s}$ is indeed in the range
predicted by the Standard Model, it should be measured soon at the 
Tevatron collider, which then determines the top-quark side of 
the unitarity triangle much more accurately. One also expects that 
the direct measurement of $\sin(2\beta)$ will soon become one of 
the most stringent constraints on $(\bar\rho,\bar\eta)$. Further 
information will come from non-leptonic $B$ decays. These two topics 
are discussed below.

If inconsistencies between the various quantities should arise, it
will be important to identify the culprit. The determinations of 
$|V_{ub}|$ from semileptonic decays and the phase $-\gamma$ of 
$V_{ub}$ from non-leptonic decays with interference of 
$b\to c\bar u D$ (no weak phase) and $b\to u\bar c D$ (weak phase $\gamma$) 
tree transitions and their conjugates ($D=d,s$) are unlikely to 
be modified strongly by new flavour-changing interactions and 
hence determine $(\bar\rho,\bar\eta)$ even in the presence of ``New 
Physics''. In contrast, most of our current knowledge 
on $(\bar\rho,\bar\eta)$ derives from meson mixing ($\epsilon$, 
$\Delta M_{B_{d,s}}$ and the direct measurement of $\sin(2\beta)$), 
which being second order in weak interactions is expected to be 
more sensitive to New Physics than non-leptonic decays. The current 
consistency of the quantities related to mixing is therefore even more 
auspicious. As already mentioned, by the end of this decade the 
unitarity triangle could be accurately determined from rare 
kaon decays alone, and also from $B$ decays. Altogether, the 
Kobayashi-Maskawa mechanism will be decisively and precisely tested.

\section{\boldmath Interpretation of $\sin(2\beta)$}

In 2001 CP violation has been observed for the first time 
in $B$ meson decays, 
more precisely in the interference of mixing and decay. Assume that 
both, $B^0$ and $\bar B^0$, can decay into a CP eigenstate $f$, 
call the amplitude of the former decay $A$, the latter $\bar A$ and 
define $\lambda=e^{-i\phi_d} \bar A/A$ with $\phi_d=2\beta$ the phase of the 
$B\bar B$ mixing amplitude in the Standard Model 
(standard phase convention). A $B$ meson 
identified as $B^0$ at time $t=0$ can decay into $f$ at a later 
time $t$ either directly or indirectly through its $\bar B^0$
component acquired by mixing. 
If there is CP violation, the amplitude for the 
CP conjugate process will be different, resulting in a time-dependent 
asymmetry
\begin{eqnarray}
\label{atime}
A_{\rm CP}(t) &=& \frac{\Gamma(\bar B^0(t)\to f)-\Gamma(B^0(t)\to f)}
{\Gamma(\bar B^0(t)\to f)+\Gamma(B^0(t)\to f)}.\nonumber\\
&&\hspace*{-1.2cm}
= \frac{2{\rm Im}\lambda}{1+|\lambda|^2}\sin(\Delta M_B t)-
\frac{1-|\lambda|^2}{1+|\lambda|^2}\cos(\Delta M_B t)
\end{eqnarray}
In deriving this result it is assumed that the lifetime difference 
of the $B$ meson eigenstates is negligible, which is a very good
approximation for $B_d$ mesons. One also assumes that CP violation 
in mixing (the corresponding $\epsilon$-parameter for $B$ mesons) 
is negligible. This has been verified at the percent level through 
the charge asymmetry in semi-leptonic decays. Alternatively, 
$\epsilon_B$ can be obtained from the time-dependent totally inclusive
decay asymmetry \cite{Beneke:1996hv}.
When $A$ is dominated by a single weak phase, 
$A=|A| e^{i\delta_W}$ (so that $|\lambda|=1$), the time-dependent 
asymmetry (\ref{atime}) is 
proportional to $\mp\sin 2(\beta+\delta_W)$, the sign depending on the 
CP eigenvalue of $f$. 

The final state $J/\psi K_S$ (and related ones) is unique in this 
respect, since the second term in the decay amplitude 
\begin{equation}
A(\bar B\to J/\psi K) = V_{cb} V_{cs}^* (T-P) + V_{ub} V_{us}^* (T_u-P)
\end{equation}
is only of the order of a percent, since it is suppressed by
$\lambda^2$ due to a small CKM factor and further suppressed 
by a penguin loop or $u\bar u\to c\bar c$ rescattering. Furthermore 
$\delta_W\approx 0$ for $b\to c\bar c s$. Hence the mixing-induced CP 
asymmetry in 
$B\to J/\psi K$ decay determines the $B\bar B$ mixing phase (relative to 
$b\to c\bar c s$), i.e. $\sin(2\beta)$ in the Standard Model, 
with little theoretical uncertainty
\cite{Bigi:1981qs,Dunietz:1986vi}. 
It determines the $B\bar B$ mixing phase also 
beyond the Standard Model, since it is unlikely that the CKM-favoured 
$b\to c\bar c s$ transition acquires a large CP-violating phase from new 
flavour-changing interactions. The assumption that the amplitude has 
only one dominant term can be partially checked by fitting a 
$\cos(\Delta M_B t)$-term to the time-dependent asymmetry and by
searching for a CP-violating asymmetry in the charged $B^+\to J/\psi
K^+$ decay and its CP-conjugate. 

The asymmetry is now precisely measured by the two $B$ factories. The
central values reported by both experiments have been increasing over the 
past year as the statistics of the experiments improved and now 
reads \cite{Aubert:2001nu,Abe:2001xe}
\begin{eqnarray}
\sin(2\beta) =
\left\{\begin{array}{ll}
0.59\pm 0.15 & \quad\mbox{BaBar} \\[0.0cm]
0.99\pm 0.15 & \quad\mbox{Belle},
\end{array}\right.
\end{eqnarray}
yielding the world average\footnote{This result has been updated by 
the new Belle data published shortly after the conference. At the time
of the conference the Belle result was 
$\sin(2\beta)=0.58^{+0.32+0.09}_{-0.34-0.10}$ resulting in the world
average $\sin(2\beta)=0.61\pm 0.13$.} $\sin(2\beta)=0.79\pm 0.10$. 
The fact that 
this asymmetry is large and in agreement with the indirect determination 
of the angle $\beta$ in (\ref{angles}) 
leads to two important conclusions on the 
nature of CP violation:
\begin{itemize}
\item CP is not an approximate symmetry of Nature (as could have been 
if CP violation in kaon decays were caused by some non-standard 
interaction).
\item the Kobayashi-Maskawa mechanism of CP violation is most likely 
the dominant source of CP violation at the electroweak scale.
\end{itemize}

The $B\bar B$ mixing matrix element is given by $M_{12}-\Gamma_{12}/2$, where 
$\Gamma_{12}$ is related to the lifetime difference and can be
neglected and $M_{12}$, including a potential contribution from 
new flavour-changing interactions, can be written as 
\begin{equation}
M_{12} = (V_{td}^* V_{tb})^2 \hat M_{12} + M_{\rm NP}.
\end{equation}
The non-standard contribution to 
$B\bar B$ mixing is constrained by the requirement that 
no conflict arises with the measurement of $\Delta M_{B_d}$ related to
the modulus of $M_{12}$. There are two generic options for the
non-standard contribution. The first is to assume that new 
flavour-violating interactions are still 
proportional to the CKM matrix so that 
$ M_{\rm NP}$ is proportional to $(V_{td}^* V_{tb})^2$. In this 
case the time-dependent CP asymmetry in $B_d\to J/\psi K$ decay
continues to determine $\sin(2\beta)$. The corresponding 
class of models is often referred 
to as ``minimal flavour violation models'' \cite{Ciuchini:1998xy}. 
If one further assumes that the effective $|\Delta B|=2$ interaction 
continues to be lefthanded, there is a strong correlation between 
modifications of $B\bar B$ mixing 
and $K\bar K$ mixing. One then finds 
that only small modifications of the $B\bar B$ mixing 
phase are possible given the constraints on 
$K\bar K$ mixing and $\Delta M_{B_{d,s}}$,  
in particular $\sin(2\beta)>0.42$ \cite{Buras:2000xq,Buras:2001af},  
and a preferred range from 
0.5 to 0.8 \cite{Bergmann:2001pm}. An extension of this analysis 
which allows a modification of $\Delta M_{B_d}/\Delta M_{B_s}$ has 
been considered in \cite{Ali:2001ej,lunghi}. The second option relaxes the 
assumption that 
$ M_{\rm NP}$ is proportional to $(V_{td}^* V_{tb})^2$, so 
that the time-dependent CP asymmetry in $B_d\to J/\psi K$ decay
is no longer directly related to $\sin(2\beta)$. In general 
the new flavour-violating interactions then contain new CP phases 
and it is possible to arrange them such that the time-dependent CP 
asymmetry in $B_d\to J/\psi K$ decay 
can take any value. There has been much interest prior to this 
conference in exploring models that allow the asymmetry to be small, 
motivated in particular by the BaBar 
measurement $\sin(2\beta)=0.12\pm 0.38$ as of summer 2000, 
which has now been superseded by the result quoted above. Generic
features of models with new CP- and flavour-violating interactions 
will be briefly described later, but the current status of the 
measurement does no longer 
mandate a detailed discussion of the ``small-$\sin(2\beta)$''
scenario. 

\section{\boldmath CP violation in $B$ meson decays}

The Kobayashi-Maskawa mechanism predicts large 
CP-violating 
effects only in $B$ decays and very rare $K$ decays. The primary 
focus of the coming years will be to verify the many 
relations between 
different observables predicted in the Kobayashi-Maskawa scenario, 
since the CKM matrix contains only a single phase.

An example of this type is the decay $\bar B_d\to\phi K$ due to 
the penguin $b\to s\bar s s$ transition at the quark level. In 
the Standard Model the time-dependent CP asymmetry of this 
decay is also proportional to $\sin(2\beta)$ to reasonable 
(though not as good) precision. However, new interactions 
are more likely to affect the loop-induced penguin transition 
than the tree decay $b\to c\bar c s$ and may be revealed 
if the time-dependent asymmetry in $\bar B_d\to\phi K$ turns out 
to be different from that in $\bar B_d\to J/\psi K$. However, 
if the difference is small, its interpretation requires that 
one controls the strong interaction effects connected 
with the presence of a small up-quark penguin 
amplitude with a different weak phase. This difficulty is 
of a very general nature in $B$ decays.

\subsection{CP violation in decay}

The need to control strong interaction effects is closely related 
to the possibility of observing CP violation in the decay 
amplitude (``direct CP violation''). 
The decay amplitude has to have at least two components 
with different CP-violating (``weak'') phases,
\begin{equation}
A(B\to f) = A_1 e^{i\delta_{S1}} e^{i\delta_{W1}}+ 
A_2 e^{i\delta_{S2}} e^{i\delta_{W2}}.
\end{equation}
If the CP-conserving (``strong interaction'') 
phases are also different, the 
partial width of the decay differs from that of its 
CP-conjugate, $\Gamma(B\to f)\not=\Gamma(\bar B\to \bar f)$. 
Many rare $B$ decays are expected to exhibit CP violation 
in decay, because a given final state can often be reached 
by different operators ${\cal O}_i$ with different CKM factors 
from the weak effective Hamiltonian 
(``tree'', ``QCD, electroweak, and magnetic penguins'') 
leading to the interference of a tree and a sizeable or 
even dominant penguin amplitude. The weak phase difference 
can only be determined, however, if the strong interaction 
amplitudes are known. This is also necessary for mixing-induced 
CP asymmetries, if the decay amplitude is not dominated by a 
single term. On the technical level, the matrix elements 
$\langle M_1 M_2|{\cal O}_i|\bar B\rangle$ have to be known for 
a two-body final state. The problem is analogous to the 
computation of $\epsilon'/\epsilon$ in kaon decay, except that 
now the initial state is heavy.

There exist two complementary approaches to obtain the strong 
interaction amplitudes, 
$\langle M_1 M_2|{\cal O}_i|\bar B\rangle$. The first, 
``traditional'' approach employs a general 
parameterization of the decay amplitudes of a set of related decays, 
implementing SU(2)-isospin relations. The remaining strong 
interaction parameters are then determined from data (often 
also using SU(3) flavour symmetry and ``little'' further 
assumptions on the magnitudes of some amplitudes). Very often this 
requires difficult measurements. The second approach attempts to 
calculate the strong interaction amplitudes directly from QCD 
with factorization methods also used in high-energy strong interaction 
processes. A systematic formulation of this approach 
has been given only recently \cite{Beneke:1999br} and makes 
essential use of the fact that 
the $b$ quark mass is large. There is currently no theoretical 
framework that also covers $1/m_b$-corrections systematically, so 
there is an intrinsic limitation to the accuracy that one can 
expect from this approach. In the following the ``QCD factorization 
approach'' to non-leptonic $B$ decays is briefly described.

\subsection{QCD factorization}

In the heavy quark limit the $b$ quark decays into very energetic 
quarks (and gluons), 
which must recombine to form two mesons. Using methods from the 
heavy quark expansion and soft-collinear factorization 
(``colour-transparency'') one can argue that 
the amplitude of a decay into two light mesons assumes 
a factorized form \cite{Beneke:1999br,Beneke:2000ry}. 
Schematically,
\begin{eqnarray}
\label{fact}
A(\bar B\to M_1 M_2) &=& F^{B\to M_1}(0)\int_0^1 \!\!du\,T^I(u)
\Phi_{M_2}(u) \nonumber\\[0.0cm]
&&\hspace*{-2.8cm}
+\!\int \!\!d\xi du dv \,T^{II}(\xi,u,v)\,\Phi_B(\xi)\Phi_{M_1}(v) 
\Phi_{M_2}(u),
\end{eqnarray}
where $F^{B\to M_1}$ is a form factor, $\Phi_X$ denote light-cone 
distribution amplitudes and $T^{I,II}$ are hard-scattering 
kernels, which can be computed in perturbation theory. ($M_1$ is 
the meson that picks up the spectator quark from the $B$ meson.) 
This result extends the Brodsky-Lepage approach to exclusive hard processes 
\cite{Lepage:1980fj}, because it shows factorization also in the 
presence of soft interactions in the $B\to M_1$ form 
factor. (There exist a number of calculations of $B$ 
decays based on the (controversial) assumption that 
the form factor is calculable in the Brodsky-Lepage approach 
\cite{Ward:1993kz,Li:1994iu,Dahm:1995ne,Keum:2000wi}.) The formula 
above implies:
\begin{itemize}
\item[-] There is no long-distance interaction between the
  constituents of the meson $M_2$ and the $(B M_1)$ system. This is 
the precise meaning of factorization.
\item[-] The second line represents a hard-gluon interaction with 
the spectator quark and appears only at order $\alpha_s$. At lowest 
order $T^I(u)$ is a constant proportional to the decay constant of 
$M_2$, so that $A(\bar B\to M_1 M_2) \propto i f_{M_2} F^{B\to
  M_1}(0)$ at lowest order. This reproduces 
``naive factorization'' \cite{Wirbel:1985ji}, 
but (\ref{fact}) implies that radiative corrections to this result can 
be computed.
\item[-] Final state rescattering is included in the hard-scattering 
kernels and therefore computable in the heavy-quark limit. In the 
heavy quark limit inelastic rescattering dominates and the sum of 
all rescatterings is dual to the partonic calculation.
\end{itemize}
The factorized form of (\ref{fact}) 
is valid up to $1/m_b$ corrections, some of which can be 
large. The extent to which the QCD factorization formalism can be 
of quantitative use is not yet fully known. It has been 
applied so far to a number of charmless two-body final states of 
pseudoscalar \cite{Beneke:2001ev,Muta:2000ti,Du:2001hr} and vector 
mesons \cite{Yang:2000xn,Cheng:2001aa} as discussed in part at 
this conference \cite{mat,Cheng:2001hk}. The QCD factorization approach 
has been 
successful in explaining the universality of strong-interaction effects 
in class-I $B\to D$ + light meson 
decays and in understanding the non-universality 
in the corresponding class-II decays
\cite{Beneke:2000ry,Neubert:2001sj}.\footnote{ 
Non-leptonic final states with $D$ mesons are reviewed 
at this conference in \cite{Ligeti:2001an}.}
It also appears to account 
naturally for the magnitude of the $\pi K$ branching fractions, 
sometimes considered as unexpectedly large, but there is currently 
no test that would allow one to conclude that the computation of 
strong interaction phases which are either of order $\alpha_s$ or 
$1/m_b$ is reliable in the case of penguin-dominated final states 
\cite{Beneke:2001ev}. 
Such tests will be possible soon. The non-observation of direct 
CP violation at the current level of sensitivity  
\cite{HameldeMonchenault:2001ae} supports 
the idea that strong rescattering effects are suppressed.  

\subsection{The angle {\boldmath $\alpha$}}

The angle $\alpha=\pi-\beta-\gamma$ can be  obtained from direct CP 
violation in decays with interference of $b\to u\bar{u} d$ (tree, $\gamma$) 
and $b\to d\bar q q$ (penguin, $\beta$) or from the mixing-induced
asymmetry in decays based on $b\to u\bar u d$. If one takes the point 
of view that the $B\bar B$ mixing phase is determined experimentally 
by the mixing-induced CP asymmetry in $B_d\to J/\psi K$ decay
irrespective of whether it is correctly described by the Standard
Model, then the second method actually determines $\gamma$.

The time-dependent CP asymmetry in $B\to \pi^+\pi^-$ decay is an example of
this type and determines $\alpha$ (or $\gamma$, depending on the point
of view), but only if the penguin amplitude is
neglected. This is now known not to be a good approximation. 
Neglecting only electroweak penguin amplitudes, the isospin amplitude system 
for the three, charged and neutral, 
$\pi\pi$ final states contains five real 
strong interaction parameters, just as many as there are independent 
branching fractions under the same assumption. Adding the
time-dependent asymmetry gives a sixth observable that allows one to 
determine $\alpha$ up to discrete ambiguities \cite{Gronau:1990ka}. 
Since this method requires 
a measurement of the small $B\to\pi^0\pi^0$ branching fractions, it has 
practical difficulties. Already bounds on the CP-averaged $\pi^0\pi^0$
branching fraction can be useful to constrain the amplitude 
system \cite{Grossman:1998jr,Charles:1999qx,Gronau:2001ff}. 
If $\mbox{Br}(\pi^0\pi^0)$ is small, the strong phase of the 
penguin-to-tree ratio cannot be large. In fact
$\mbox{Br}(\pi^0\pi^0)= 0$ implies $\mbox{Br}(\pi^+\pi^-)= 2\,
\mbox{Br}(\pi^\pm\pi^0)$. Conversely, a deviation from the last 
relation implies that $\mbox{Br}(\pi^0\pi^0)$ cannot be too small. 
Further constraints on the $\pi\pi$ 
modes can be obtained only by assuming also SU(3) or U-spin
symmetry. This relates, for example, 
 $B_d\to\pi^+\pi^-$ to $B_s\to K^+ K^-$. The 
inverted CKM hierarchy of penguin and tree amplitude in 
the second decay can in principle 
be used to determine $\gamma$ from a combined
measurement of the time-dependent and direct CP 
asymmetries in both decays \cite{Fleischer:1999pa}. 

Other methods have been devised that allow one to eliminate all
hadronic parameters by measurements without the need to measure the 
$\pi^0\pi^0$ mode. One method uses the interference of CP-violating 
phases with CP-conserving phases from the resonant $\rho$-meson 
propagator in the decays 
$B_d\to \{\rho^+\pi^-,\rho^0\pi^0,\rho^-\pi^+\}\to \pi^+\pi^-\pi^0$ 
\cite{Snyder:1993mx,Quinn:2000by}. The disadvantage of this 
method is that the analysis of the Dalitz plot 
of the three-particle final state requires comparatively large 
statistics. In addition, theoretical difficulties due to 
non-resonant pion production have not yet been fully removed.

\begin{figure}[t]
\epsfxsize=6cm
\hspace*{0.7cm}
\epsffile{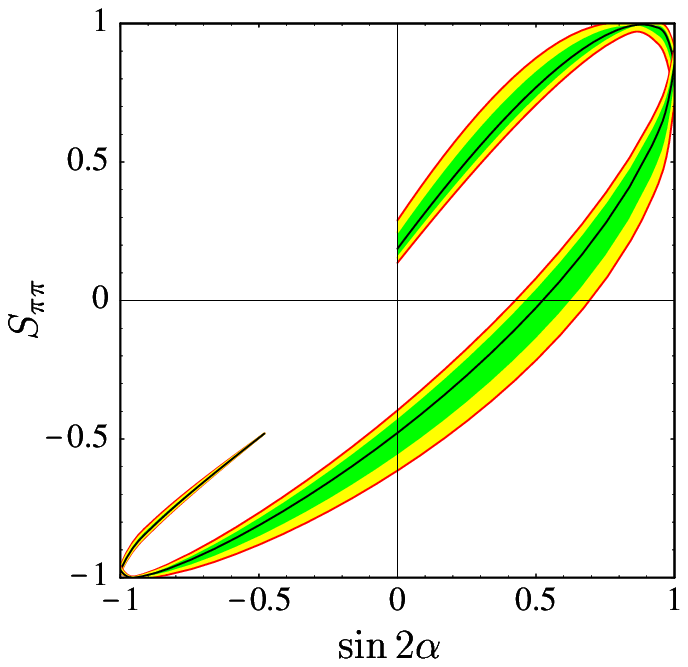}
\epsfxsize=6.4cm
\vskip-6cm
\hspace*{3.2cm}
\centerline{\epsffile{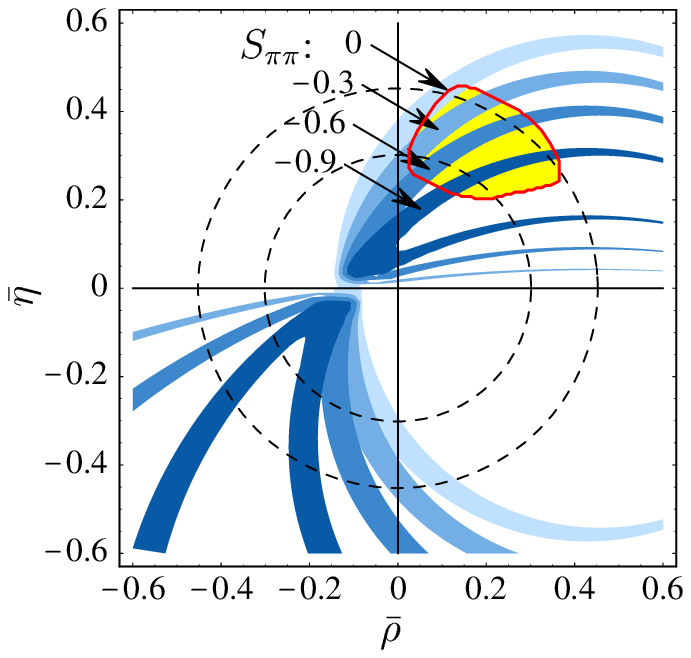}}
\caption[dummy]{\label{fig3} Correlation between the CP-asymmetry 
and $\sin(2\alpha)$ (left) and the corresponding constraint in the 
$(\bar\rho,\bar\eta)$ plane (right) \cite{Beneke:2001ev}.}
\end{figure}

The angle $\alpha$ can be determined from the 
time-dependent CP asymmetry in $B\to \pi^+\pi^-$ decay alone, if 
the relative magnitude of the penguin amplitude, $P/T$, can be computed. 
The asymmetry (\ref{atime}) is given by
\begin{equation}
   A_{\rm CP}[\pi\pi](t) =  S_{\pi\pi} \sin(\Delta M_{B_d}\,t)
   + A_{\rm CP}^{\rm dir}[\pi\pi] \cos(\Delta M_{B_d}\,t),
\end{equation}
where $S_{\pi\pi} = \sin(2\alpha)$, if $P/T=0$, in which case 
the direct CP asymmetry $A_{\rm CP}^{\rm dir}[\pi\pi]$ vanishes. 
The direct CP asymmetry is proportional to the sine of the strong 
phase of $P/T$ and can be used as a phenomenological check of the
computation of $P/T$. Figure~\ref{fig3} displays the correlation 
between the CP-asymmetry $S_{\pi\pi}$ 
and $\sin(2\alpha)$ (left) and the corresponding constraint in the 
$(\bar\rho,\bar\eta)$ plane (right), when $P/T$ is 
computed in the QCD factorization approach \cite{Beneke:2001ev}. 
The Figure illustrates that even if theoretical (or experimental) 
uncertainties prevent an accurate determination of $\sin(2\alpha)$ 
in this way, the inaccurate result on $\sin (2\alpha)$ still translates into 
a useful constraint in the $(\bar\rho,\bar\eta)$ plane. This reflects 
the fact that other observables do not constrain $\sin(2\alpha)$ very well 
as seen from (\ref{angles}).

%%%%%%%%%%%%%%%%%%%%%%%%%%%%%%%%%%%%%%%%%%%%%%%%%%%%%%%%%%%%%%%%%%%
%%%%%%%%%%%%%%%%%%%%%%%%%%%%%%%%%%%%%%%%%%%%%%%%%%%%%%%%%%%%%%%%%%%

\subsection{The angle {\boldmath $\gamma$}}

The preferred methods to determine directly the angle $\gamma$, 
the phase of $V_{ub}^*$, rely on decays with interference of 
$b\to c\bar u D$ (no phase) and $b\to u\bar c D$ (phase $\gamma$) 
transitions and their conjugates ($D=d,s$). These decays receive no penguin 
contributions and are arguably insensitive to new flavour-changing 
interactions. $\gamma$ can be extracted from either of the 
following decay classes, $B_d(t)\to D^\pm \pi^\mp$
\cite{Dunietz:1987bv}
(or more recent variants 
\cite{Diehl:2001ey}), 
$B_d(t)\to D K_S$ \cite{Gronau:1990ra}, $B^\pm \to K^\pm D_{\rm CP}$ 
\cite{Gronau:1991dp}, $B_s(t)\to D_s^\pm K^\mp$ 
\cite{Aleksan:1991nh}, 
since every one of them provides sufficiently many observables to eliminate 
all strong interaction parameters. 
None of these strategies is simple to carry out experimentally, 
however, since they involve either small CP asymmetries, or 
small branching fractions, or disparate amplitudes, or 
rapid $B_s$ oscillations.

The possibility to determine $\gamma$ from decays with interference 
of $b\to u\bar u D$ (tree, phase $\gamma$) and $b\to Dq\bar q$ 
(penguin, phase 0 ($D=s$), $\beta$ ($D=d$)) transitions has 
therefore been thoroughly investigated recently, in particular 
the decays $B\to \pi K$. The branching fractions for these modes 
are of order $10^{-5}$ and have already been measured with an error of 
$\pm (10-20)\%$ \cite{HameldeMonchenault:2001ae}, 
including first measurements of direct CP 
asymmetries (all compatible with zero). The drawback of these and related 
modes is that the amplitudes contain more strong interaction parameters 
than there are 
observables. SU(3) symmetry and the structure of the weak effective 
Hamiltonian allow one to construct a number of interesting bounds 
on $\gamma$ \cite{Fleischer:1998um,Neubert:1998jq}, 
but a full understanding of these modes requires a 
calculation of the penguin-to-tree amplitude ratio including its 
strong rescattering phase. 

The $B\to \pi K$ decays are penguin-dominated, because the tree 
amplitudes are CKM suppressed. The final states $\pi^0 K^\pm$ and 
$\pi^0 K^0$ have significant electroweak penguin contributions. The
amplitude system contains 11 real strong interaction parameters.
Flavour symmetry is useful
to constrain some of the amplitude parameters:
\begin{itemize}
\item[-] Isospin symmetry implies \cite{Neubert:1999re}
\begin{equation}
{\displaystyle \mbox{Br}(\pi^0 \bar K^0) = 
\frac{\mbox{Br}(\pi^+ K^-)\mbox{Br}(\pi^- \bar K^0)}
{4 \mbox{Br}(\pi^0 K^-)}}\times\left\{1+O(\epsilon^2)\right\},
\end{equation}
where $\epsilon\sim 0.3$ is related to the tree-to-penguin 
ratio.\footnote{Here and in the remainder of this section ``Br'' 
always refers to branching fractions averaged over a decay mode and
its CP-conjugate.} 
Unless the correction term is unexpectedly large this relation 
suggests a $\pi^0 K^0$ branching fraction of order $6\times 10^{-6}$, 
about a factor $1.5-2$ smaller than the current measurements. 
\item[-] SU(3) or U-spin symmetry imply:
\begin{itemize}
\item[a.] The dominant electroweak penguin amplitude is 
determined \cite{Neubert:1998jq}.
\item[b.] The magnitude of the tree amplitude for $I=3/2$ final states
  is related to $\mbox{Br}(\pi^\pm\pi^0)$.
\item[c.] Rescattering and annihilation contributions to 
the (otherwise) pure penguin decay $B^+\to\pi^+K^0$ are 
constrained by  $\mbox{Br}(K^+ K^0)$, where they are CKM enhanced
relative to the penguin amplitude \cite{Falk:1998wc}.
\end{itemize}
\end{itemize}
SU(3) flavour symmetry together with a few further dynamical
assumptions (detailed below) suffice to derive bounds on $\gamma$ 
from CP-averaged branching fractions alone. The 
inequality \cite{Fleischer:1998um}
\begin{equation}
\sin^2\gamma \leq {\displaystyle \frac{\tau(B^+)}{\tau(B_d)}\,
\frac{\mbox{Br}(\pi^+ K^-)}{\mbox{Br}(\pi^-\bar{K}^0)}} \equiv R
\end{equation}
excludes $\gamma$ near $90^\circ$ if $R<1$ and is derived upon
assuming that the rescattering contribution mentioned above and 
a colour-suppressed electroweak penguin amplitude are negligible. 
Current data give $R=1.06\pm 0.18$. The ratio of charged decay 
modes satisfies \cite{Neubert:1998jq} 
(neglecting again the rescattering contribution to 
$B^+\to\pi^+K^0$ which appears here suppressed by a factor 
$\bar \epsilon_{3/2}$) 
\begin{equation} 
\label{nrbound}
2\cdot {\displaystyle
\frac{\mbox{Br}(\pi^0 K^-)}{\mbox{Br}(\pi^-\bar{K}^0)}} \equiv
R_\ast^{-1} \leq \left(1+\bar{\epsilon}_{3/2} 
\left|q-\cos\gamma\right|\right)^2 
+ \,\bar{\epsilon}_{3/2}^2\sin^2\gamma,
\end{equation}
where $q$ and $\bar \epsilon_{3/2}$ are determined according to a. and b. 
above, respectively. This bound is particularly interesting, since, 
if $R_\ast^{-1}>1$,  
it excludes a region in $\gamma$ around $55^\circ$, which is favoured 
by the indirect unitarity triangle constraints (\ref{angles}). 
Current data give 
$R_\ast^{-1}=1.40\pm 0.23$. This prefers $\gamma>90^\circ$, but the 
error is still too large to speculate about the implications of this 
statement.  A similar reasoning 
applies to the final states $\pi^+ K^-$, $\pi^0 K^0$ and their 
CP-conjugates in the decay of the neutral $B_d$ and $\bar B_d$ 
mesons \cite{Buras:1999rb},
for which the time-dependent CP-asymmetry $B_d(t)\to \pi^0 K_S$
provides an additional observable that could be
used to constrain the system of hadronic quantities.
Eq.~(\ref{nrbound}) can be turned into a determination 
of $\gamma$ if one assumes that the strong phase of the tree amplitude
relative to the penguin amplitude is not too
large \cite{Beneke:2001ev}. This assumption is justified by theoretical 
calculations as discussed next, 
but will eventually be verified experimentally by the 
observation of small direct CP asymmetries.

\begin{figure}[t]
\vspace*{0.2cm}
\epsfxsize=9cm
\centerline{\epsffile{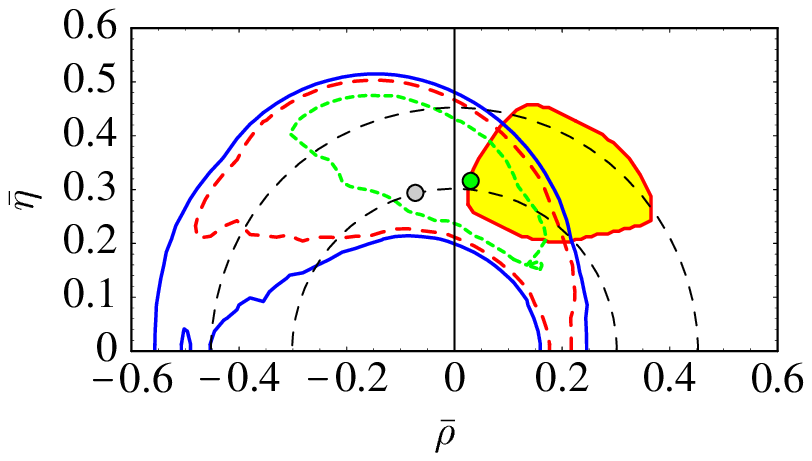}}
%\vspace*{-0.6cm}
\caption{\label{fig4}
95\% (solid), 90\% (dashed) and 68\% (short-dashed) confidence level 
contours in the $(\bar\rho,\bar\eta)$ plane obtained from a global 
fit to the CP averaged $B\to\pi K,\pi\pi$ branching fractions, using 
the scanning method as described in 
\cite{Hocker:2001xe}. The right dot shows the
overall best fit, whereas the left dot indicates the best fit for the
default hadronic parameter set. The light-shaded region indicates the region 
preferred by the standard global fit, excluding the direct measurement 
of $\sin(2\beta)$.}
\end{figure}

The possibility to compute strong interaction effects in non-leptonic 
$B\to \pi K,\pi\pi$ decays with the QCD factorization method 
and to determine $\gamma$ has been investigated in detail 
\cite{Beneke:2001ev}. 
Figure~\ref{fig4} shows the result of a global fit of $(\bar \rho,
\bar\eta)$ to CP-averaged $B\to\pi K,\pi\pi$ branching fractions. 
The result is 
consistent with the standard fit (overlaid light-shaded region) 
based on meson mixing and $|V_{ub}|$, 
but shows a preference for larger $\gamma$ or smaller $|V_{ub}|$. 
If the estimate of the theory uncertainty (included in the curves 
in the Figure) is correct, non-leptonic decays together with 
$|V_{ub}|$ from semileptonic decays already imply the existence 
of a CP-violating phase of $V_{ub}$ at the 2-3 $\sigma$ level. 
Similar conclusions have been 
obtained in \cite{He:1999mn,He:2000ys} 
with different theory inputs and no attempt to quantify 
the theoretical error. On the other hand
the analysis in \cite{Ciuchini:2001pk} sacrifices some theoretical 
input and enlarges the fit by the corresponding hadronic parameter 
to conclude that no determination of $\gamma$ is possible from 
$B\to\pi K,\pi\pi$ decays. 

%%%%%%%%%%%%%%%%%%%%%%%%%%%%%%%%%%%%%%%%%%%%%%%%%%%%%%%%%%%%%%%%%%%
%%%%%%%%%%%%%%%%%%%%%%%%%%%%%%%%%%%%%%%%%%%%%%%%%%%%%%%%%%%%%%%%%%%

\section{\boldmath Beyond $B$ and $K$ decays}

CP violation may occur outside the $K$ and $B$ meson systems, but the
pattern of flavour-changing interactions implied by the CKM matrix
leads to the conclusion that only null effects are expected in the
Standard Model at the current levels of experimental sensitivity. 
CP-violating observables in $D$ meson decays and flavour-conserving 
reactions could therefore provide unambiguous evidence for extensions
of the Standard Model which contain new flavour-changing interactions
and new sources of CP violation. Here is a brief list of topics which 
had to be omitted in this talk:\footnote{
The experimental and theoretical status of $D\bar D$ mixing 
has been  reviewed in Ligeti's talk \cite{Ligeti:2001an}.}

{\em Charmed mesons.} $D\bar D$ mixing is strongly suppressed by the 
GIM mechanism. CP-violating 
phenomena are further suppressed by small CKM factors, so that 
CP violation is small in mixing and direct CP asymmetries in charm 
decays are expected to be at most at the permille level. From a
theoretical point of view the properties of charm mesons are
especially difficult to compute reliably, since neither chiral
perturbation theory nor the heavy quark limit are useful. The charm 
system is good for order-of-magnitude effects in extensions of the 
Standard Model and the experimental bounds are beginning to approach
an interesting region for such effects.

{\em Flavour-conserving CP violation.} CP violation without flavour
violation is closely related to the strong CP problem in the Standard
Model. Once $\tilde\theta$ is set to zero by {\em fiat},
flavour-conserving CP violation induced by the CKM matrix is 
unobservably small. The situation is very different in extensions of 
the Standard Model, which can contain 
flavour-conserving CP-violating interactions at tree level, for
instance in the scalar potential. For this reason bounds on electric 
dipole moments of the neutron and the leptons put very important 
constraints on such extensions. At future colliders one can search for
CP violation in top quark and Higgs interactions in high energy 
collisions. Interesting signals are only expected in extensions of the
Standard Model. These topics are reviewed 
in \cite{Bernreuther:1998ju}.

{\em CPT violation.} CPT has been assumed to be a good symmetry in
this talk and no distinction between CP and T violation has been made. As
is well-known, CPT symmetry is a general consequence of locality and
relativistic invariance in quantum field theories. 
Insisting on locality of the effective field
theory, CPT violation is most naturally discussed as a consequence of 
broken Poincar\'{e} invariance. At this 
conference consequences of CPT non-invariance in the evolution 
and decay of entangled meson-antimeson states \cite{Bernabeu:2001xg}
and the possibility of an anomalous CPT symmetry in theories with 
chiral fermions \cite{Klinkhamer:1999zh} have been presented.

\section{CP violation in extensions of the Standard Model}

The emerging success of the Kobayashi-Maskawa mechanism of 
CP violation is sometimes accompanied by a sentiment of disappointment 
that the Standard Model has not finally given way to a more fundamental 
theory. The implications of this success are, perhaps, more 
appreciated, when it is viewed from the perspective of the year 1973, 
when the mechanism was conceived.  
After all, the Kobayashi-Maskawa mechanism predicted a new 
generation of particles on the basis of the tiny and obscure effect 
of CP violation in $K\bar K$ mixing. It then predicted relations between 
CP-violating quantities in $K$, $D$, $B$-physics which {\em a priori} 
might be very different. The fact that it has taken nearly 30 years 
to assemble the experimental tools to test this framework does 
not diminish the spectacular fact that once again Nature has realized 
a structure that originated from pure reasoning.

Nevertheless several arguments make it plausible that the 
Kobayashi-Maskawa mechanism is not the final word on CP violation. 
The strong and cosmological CP problem (baryogenesis) continue to 
call for an explanation, probably related to high energy scales. 
There may be an aesthetic appeal to realizing the full Poincar\'{e} 
group as a symmetry of the Lagrangian, in which case CP and P 
symmetry breaking must be spontaneous. One of the strongest arguments 
is, however, that the electroweak hierarchy problem seems to 
require an extension of the Standard Model at the TeV scale. 
Generic extensions have more sources of CP violation than the 
CKM matrix. These have not (yet) been seen, suggesting that there 
is some unknown principle that singles out the CKM matrix as 
the dominant source of flavour and CP violation. In the following 
I give a rather colloquial overview of CP violation in generic 
extensions of the Standard Model at the TeV scale. This is perhaps an 
academic catalogue, but it illustrates how restrictive the 
Kobayashi-Maskawa framework is.

\subsection{Extended Higgs sector}

Extending the Higgs sector by just a second doublet opens many 
new possibilities. The Higgs potential may now contain complex 
couplings, leading to Higgs bosons without definite CP parity, 
to CP violation in charged Higgs interactions, flavour-changing 
neutral currents, and CP violation in flavour-conserving interactions 
such as $t\bar t H$ and electric dipole moments. 
The Lagrangian could also be CP-conserving with CP violation 
occurring spontaneously through a relative phase of the two 
Higgs vacuum expectations values \cite{Lee:iz}. 

Without further restrictions both scenarios already cause too 
much CP violation and flavour-changing neutral currents, so that 
either the Higgs bosons must be very heavy or some special 
structure imposed. For example, discrete symmetries may 
imply that up-type and down-type quarks couple to only one Higgs 
doublet, a restriction known as ``natural 
flavour conservation'' \cite{Glashow:1976nt,Paschos:1976ay}, 
since it forbids flavour-changing neutral currents (and also 
makes spontaneous CP violation impossible with only two 
doublets). With flavour conservation imposed, CP and flavour violation 
occur through the CKM matrix, but in addition to the usual 
charged currents also in charged Higgs interactions. This is 
usually considered in the context of supersymmetry, since 
extended Higgs models suffer from the same hierarchy problem 
as the Standard Model.

\subsection{Extended gauge sector (left-right symmetry)}

Left-right-symmetric theories with gauge group SU(2)${}_{\rm L}\times$SU(2)$
{}_{\rm R}\times$U(1)${}_{\rm B-L}$ are attractive
\cite{Mohapatra:1974hk}, 
because parity and CP symmetry can be broken spontaneously. The minimal model 
requires already an elaborate Higgs sector (with triplets in addition 
to doublets) and suffers from the hierarchy problem. CP violation 
in the quark sector now occurs through left- and right-handed 
charged currents with their respective CKM matrices. But since 
all CP violation arises through a single phase in a Higgs vacuum 
expectation value, there is now a conflict between suppressing 
flavour-changing currents to a phenomenologically acceptable 
level and the need to make the effects of this phase large enough to 
generate the CP violating phenomena already observed. The minimal 
left-right symmetric model is therefore no longer viable 
\cite{Ball:1999mb,Barenboim:2001vu}. 

Spontaneous CP symmetry breaking is excluded in the minimal
supersymmetric extension of the Standard Model, but it is an option, 
when the model is augmented by an extra gauge singlet superfield 
(next-to-minimal supersymmetric standard model) or by the choice of 
an enlarged Higgs sector, as discussed at this conference  
\cite{Teixeira:2001mu,Chen:2001ty}.

\subsection{Extended fermion sector}

The Standard model can be extended by an extra $d$-type quark 
with electric charge $-1/3$ \cite{delAguila:1985mk}. 
This quark should be a weak singlet in 
order not to conflict electroweak precision tests. After electroweak 
symmetry breaking, the down-quark mass matrix must be diagonalized by a
unitary $4\times 4$ matrix. The motivation for such an extension of 
the Standard Model may be less clear, in particular as there is 
no symmetry principle that would make the extra singlet-quark 
naturally light. 
However, this theory provides an example in which the unitarity ``triangle'' 
does no longer close to a triangle, but is extended to 
a quadrangle: 
\begin{eqnarray}
&&V_{ud} V_{ub}^*+V_{cd} V_{cb}^*+\underbrace{V_{td} V_{tb}^*}
\,\,\,+\,\,\,\underbrace{U_{db}}\,\,\,=0\\
&& \hspace*{2.5cm}\approx 8\cdot 10^{-3}\quad {<10^{-3}}\nonumber
\end{eqnarray}
The unitarity triangle ``deficit'' $U_{db}$ also determines the 
strength of tree-level flavour-changing $Z$ boson couplings and 
is currently constrained by $B\bar B$ mixing and rare decays to 
about a tenth of the length of a side of the triangle. (The 
corresponding coupling $U_{ds}$ is constrained much more tightly in 
the kaon system by the non-observation of $K\to \mu^+\mu^-$ decay.) 
This model could in principle still give large 
modifications of $B\bar B$ mixing and non-leptonic $B$ decays, 
including CP asymmetries \cite{Barenboim:2001fd}.

\subsection{Supersymmetry}

The minimal supersymmetric standard model 
\cite{Dimopoulos:1981zb,Sakai:1981gr}
is arguably the most natural 
solution to the electroweak hierarchy problem, but it is not particularly 
natural in its most general form from the point of view of CP violation. 
The Lagrangian including the most general renormalizable operators 
that break  
supersymmetry softly contains 44 CP-violating constants of nature. 
(This does not yet include a neutrino mass matrix and also assumes 
that R-parity is conserved.) 
One of them is the usual CKM phase which appears in charged current and 
chargino interactions. Three phases appear in flavour-conserving 
CP observables, 27 in flavour- and CP-violating quark-squark-gluino 
interactions (squark mass matrices and $A$-terms) and 13 in the 
(s)lepton sector. 

The flavour-conserving phases must be small to comply with the 
non-observation of electric dipole moments. An intriguing feature of 
supersymmetry is the existence of CP and flavour violation in 
strong interactions, that is the possibility of a flavour-changing, 
CP-violating quark-squark-gluino vertex. These interactions can be 
much stronger than the Standard Model weak interactions  
and to suppress them 
to a phenomenologically acceptable level, one has to assume 
that either (some of) the masses of superparticles are rather large, 
or that the squark mass matrices are diagonal in the same basis that 
also diagonalizes the quark mass matrices (alignment) or that 
the squarks have degenerate masses, in which case a generalization 
of the GIM mechanism suppresses flavour-changing couplings 
\cite{Ellis:1981ts,Donoghue:1983mx}. Since 
almost all CP-violating phases of the minimal supersymmetric standard model 
originate from supersymmetry breaking terms, one must understand 
supersymmetry breaking to answer the question why CP and flavour 
violation are so strongly suppressed. There exist mechanisms which can 
naturally realize one or the other of the conditions listed above 
(for example supersymmetry breaking through gauge 
interactions \cite{Giudice:1998bp}), 
but none of the mechanisms is somehow singled out. 

There is currently much activity aiming at constraining the flavour- 
and CP-violating couplings from the many pieces of data that become 
now available. In fact these couplings are so many-fold that the 
CP-violating effects observed in kaon and $B$ meson decays can all 
be ascribed to them (allowing, in particular, the CKM phase to be
small) at the price of making the consistency of the 
Kobayashi-Maskawa mechanism appear accidental. For example, 
several mechanisms, making use of flavour-changing strong
interactions, have been proposed that could enlarge $\epsilon'/\epsilon$
\cite{Colangelo:1998pm,Masiero:1999ub,Kagan:1999iq} without 
conflicting other data, and the impact of these interactions on 
$B$ decays has been discussed in two talks at this conference 
\cite{Arhrib:2001gr,Becher:2001zb}. 
The hope could be that eventually some pattern of 
restrictions on these small couplings will be seen that could give a hint 
on the origin of supersymmetry breaking. It is also plausible to 
assume that strong flavour and CP violation is absent (or too small 
to observe) in supersymmetry 
for one or the other yet unknown reason. 
Neglecting also the flavour-conserving 
CP-violating effects, the CKM matrix is then the only effect 
of interest. The presence of additional particles with CKM couplings 
still implies modifications of meson mixing and rare decays, 
but these modifications are now much smaller and, in general 
(but excepting rare radiative or leptonic decays), 
precise theoretical results are needed to disentangle them from 
hadronic uncertainties. 

Whatever the outcome of the search for new CP violation may be, 
it will restrict the options for model building severely. The current 
data point towards a privileged standing of the CKM matrix. However, 
a theoretical rationale for this privileged standing is yet to 
be discovered.

\section{Conclusions}

\noindent 
I. The (expected) observation of large CP violation 
in $B$ decays together with $\epsilon^\prime/\epsilon$ and 
the consistency of indirect determinations
of the unitarity triangle imply that:  

\begin{itemize}
\item[-] CP is not an approximate 
symmetry of Nature -- rather CP violation is rare in the Standard 
Model because of small flavour mixing.

\item[-] the Kobayashi-Maskawa mechanism of CP violation 
in charged currents is {\em probably} the {\em dominant} source of CP 
violation {\em at the electroweak scale}.
\end{itemize}

\noindent 
II. CP and electroweak symmetry breaking provide complementary 
motivations to search for extensions of the Standard Model, 
but: 
\begin{itemize}
\item[-] on the one hand, there exists 
no favoured candidate model for CP violation beyond the Standard Model -- 
rather there is a CP problem in many conventional extensions.
\item[- ]on the other hand, baryogenesis requires CP violation 
beyond the Standard Model, probably decoupled from CP violation 
observable at accelerators.
\end{itemize}

\noindent 
III. The study of CP violation is at a turning point with many new 
experimental capabilities and new theoretical methods to interpret 
non-leptonic decay data. Perhaps the most important result  
of the near future, however, will be to find (or not find) the 
$B_s$ mass difference $\Delta M_{B_s}\approx 17.5\,\mbox{ps}^{-1}$, 
confirming once more the Standard Model paradigm (or to put it 
into serious difficulty).

\section*{Acknowledgement}

\noindent
I would like to thank Gautier~Hamel de Monchenault for assistence with
experimental data, and M.~Diehl, T.~Feldmann and A.P.~Chaposvky for 
careful reading of the manuscript.

\end{document}